\documentclass[aps,groupedaddress,showpacs,showkeys,twocolumn,amsfonts]{revtex4}

\newcommand{\bfk}{\mbox{\boldmath $k$}}
\newcommand{\bfx}{\mbox{\boldmath $x$}}
\newcommand{\aalpha}{{\varphi}}
\newcommand{\abeta}{{\xi}}
\newcommand{\thalf}{{\textstyle{\frac{1}{2}}}}

\begin{document}

\title{Phase space localization of antisymmetric functions}

\author{L.L. Salcedo}
\email{salcedo@ugr.es}

\affiliation{
Departamento de F\'{\i}sica Moderna, Universidad de Granada, E-18071
Granada, Spain }

\date{\today} 

\begin{abstract}
Upper and lower bounds are written down for the minimum information
entropy in phase space of an antisymmetric wave function in any number
of dimensions. Similar bounds are given when the wave function is
restricted to belong to any of the proper subspaces of the Fourier
transform operator.
\end{abstract}



\maketitle

{\bf 1.} Heisenberg uncertainty relations between position and
momentum can be viewed as a constraint on the maximum allowable
localization of quantum states in phase space, thereby displaying a
basic principle of Quantum Mechanics. In this approach the
localization is based on the size of the state. An alternative
approach \cite{Deutsch:1983,Partovi:1983vb,Bialynicki-Birula:1985bk,%
Gadre:1985,Maassen:1988,Yanez:1994,Hall:1994,Buzek:1995,%
Sanchez-Ruiz:1995,Majernik:1997,Orlowski:1997,Massen:1998zy,Ion:1999zw}
is based on using the information entropy of the state as a measure of
its localization. To be specific, if $\psi(x)$ is a normalized wave
function, the information entropy of the corresponding state in
position space is defined as
\begin{equation}
S_x(\psi)= -\int d^d x \,|\psi(\bfx)|^2 \log(|\psi(\bfx)|^2) \,,
\end{equation}
(where $d$ denotes the dimension of the space).  The more localized
the state the lower the entropy. For instance, for a one-dimensional
Gaussian $\psi(\bfx)= (\pi a^2)^{-1/4}\exp(-x^2/2a^2)$ (with $a>0$),
the behavior of the entropy $S_x= (1+\log(\pi a^2))/2$ is
qualitatively similar to that of the dispersion $\Delta x=
a/\sqrt{2}$. However, for a state composed of two well separated
Gaussian functions of given widths, the dispersion would increase with
the separation whereas the entropy would not. This illustrates the
different concept of localization in both approaches.

The entropy in momentum space is defined similarly as
\begin{equation}
S_k(\psi)= 
-\int \frac{d^d k}{(2\pi\hbar)^d} \,|\tilde\psi(\bfk)|^2 
\log(|\tilde\psi(\bfk)|^2) \,,
\end{equation}
where
\begin{equation}
\tilde\psi(\bfk)= \int d^d x \,e^{-i\bfk\bfx/\hbar}\psi(\bfx) \,.
\end{equation}
For the one-dimensional Gaussian this gives $\tilde\psi(\bfk)=(4\pi
a^2)^{1/4}\exp(-a^2 k^2/2\hbar^2)$ and $S_k= (1-\log(4\pi a^2))/2$,
once again displaying a relation between localization in momentum
space and entropy in that space. It is then natural to introduce an 
entropy in phase space as
\begin{equation}
S(\psi)= S_x(\psi)+ S_k(\psi)\,.
\end{equation}
This definition respects invariance under dilatations (as also does
$\Delta x\Delta p$). In the entropic approach the uncertainty
principle is expressed by the fact that $S_x$ and $S_k$ cannot both be
lowered simultaneously without limit. More precisely, the entropy in
phase space satisfies the lower bound
\cite{Beckner:1975,Bialynicki-Birula:1975,Hirschman:1957}
\begin{equation}
S(\psi) \ge d(1-\log 2) \,.
\end{equation}
This bound is sharp and is attained by the Gaussian functions (with
non complex width). These are the unique minimizers
\cite{Lieb:1990}. In the one dimensional case this bound implies
Heisenberg's relation $\Delta x\Delta p \ge \hbar/2$
\cite{Hirschman:1957,Bialynicki-Birula:1975}. We note that, in the
literature, the name phase space entropy is used with various
meanings, and frequently refers to entropies of matrix densities
defined on phase space by means of a Wigner or a Husimi transformation
(Wehrl entropy) \cite{Buzek:1995}. $S$ can also be viewed as the
information entropy of a probability density in phase space, namely,
$\rho(\bfx,\bfk)=|\psi(\bfx)|^2|\tilde\psi(\bfk)|^2$.

{\bf 2.} Consider now two particles constrained to have a symmetric or
an antisymmetric relative wave function (say, two electrons in a
singlet or triplet spin state, respectively). One can ask which is the
most localized state in phase space in both cases. In the symmetric
case the answer is just a Gaussian function since this is the absolute
minimum of $S(\psi)$ and it is symmetric. In the {\em antisymmetric}
case the solution is not known but certainly there exists a sharp
lower bound, which will be denoted $C_d$
\begin{equation}
S(\psi) \ge C_d \ge d(1-\log 2)\,,\qquad \psi(\bfx)=-\psi(-\bfx) \,.
\end{equation}
That the bound $C_d$ is sharp implies that there exists a (minimizing)
sequence of antisymmetric functions $\psi_n$ such that
$\lim_{n\to\infty} S(\psi_n)=C_d$. It is not guaranteed, however, that
$S(\psi)=C_d$ is fulfilled for any function $\psi(x)$ in
$L^2({\mathbb{R}}^d)$.

In view of the fact that in the symmetric case the minimizer is a
Gaussian, the ground state of a quantum harmonic oscillator, a natural
guess would be to take the first excited state for the minimizer in
the antisymmetric case. In one dimension this gives
\begin{equation}
S= -1 +\log 2 + 2\gamma \ge C_1
\end{equation}
where $\gamma$ is Euler's Gamma. Numerically $S=0.847579$. This is
not, however, the true minimizer, since in \cite{Salcedo:1998pe} a
better guess $\psi_0$ was proposed which yields
\begin{equation}
S(\psi_0)= 2(1 -\log 2) \ge C_1
\end{equation}
(numerically $S= 0.613706$). In fact, based on strong numerical
evidence, this was conjectured in \cite{Salcedo:1998pe} to be the true
sharp bound in one dimension. (We note that $\psi_0$ represents
actually a minimizing sequence and not a truly acceptable state.)

{\bf 3.} Whether this conjecture is true or not, one can now put an
upper bound on $C_d$ for arbitrary $d$. Let $\psi_0(x)$ be the
one-dimensional best guess for the antisymmetric case found in
\cite{Salcedo:1998pe} and let $g(\bfx)$ be a $(d-1)$-dimensional
Gaussian function, then for the antisymmetric $d$-dimensional function
\begin{equation}
\psi(\bfx)= \psi_0(x_1)g(x_2,\ldots,x_d)
\label{eq:1}
\end{equation}
we can immediately evaluate the entropy noting that this quantity is
additive for separable functions
\begin{equation}
S(\psi)= S(\psi_0)+S(g)\,.
\end{equation}
This implies the bounds
\begin{equation}
(d+1)(1 -\log 2) \ge C_d \ge d(1 -\log 2) \,.
\end{equation}

The Fourier transform can be regarded as an operator from
$L^p({\mathbb{R}}^d)$ into $L^q({\mathbb{R}}^d)$, where
$p^{-1}+q^{-1}=1$ and $1<p\le 2\le q$ \cite{Lieb:1997bk}. As already
noted in \cite{Hirschman:1957} the minimization of the functional $S$
in ${\mathbb{R}}^d$ is directly related to the norm of the Fourier
transform operator, defined as the supremum of
$||\tilde\psi||_q/||\psi||_p$ on ${\mathbb{R}}^d$. [This is because
derivating $\int d^dx|\psi|^p$ or $(2\pi\hbar)^{-d}\int d^dk
|\tilde\psi|^q$ with respect $q$ at $q=2$ generates $S_x(\psi)$ and
$S_k(\psi)$ respectively.] The relation between norm and entropy
extends immediately for the subspace of antisymmetric functions
\cite{Salcedo:1998pe}. Let $K_{d,q}$ be the norm of the Fourier
operator on the antisymmetric subspace of ${\mathbb{R}}^d$. For this
quantity, using the same separable function introduced in
(\ref{eq:1}), we find the bounds
\begin{equation}
(p^{1/p}q^{-1/q})^{(d+1)/2} \le K_{d,q} \le (p^{1/p}q^{-1/q})^{d/2} \,.
\end{equation}

{\bf 4.} The form of $\psi_0(x)$ is quite remarkable. It is an
antisymmetric array of very narrow Gaussian functions located at
equally spaced points and with amplitudes modulated by a very wide
Gaussian function. Namely,
\begin{equation}
\psi_0(x)= e^{-\pi a^2 x^2}\sum_{n\in \mathbb{Z}}(-1)^n e^{-\pi
(x-n-\frac{1}{2})^2/a^2}\,,\quad a\to 0^+\,.
\end{equation}
As indicated, $a$ is a vanishingly small positive parameter. (We will
implicit assume this in what follows.) Alternatively one can use
\begin{equation}
\psi_0(x)= \sum_{n\in \mathbb{Z}}(-1)^n e^{-\pi a^2
(n+\frac{1}{2})^2}e^{-\pi (x-n-\frac{1}{2})^2/a^2}\,,\quad a\to 0^+\,.
\end{equation}
Both expressions coincide for small $a$. This function is
normalized. In the sense of distributions, $\psi_0(x)/a$ is equivalent
to
\begin{equation}
d_0(x)= \sum_{n\in \mathbb{Z}}(-1)^n \delta(x-n-\thalf) \,.
\end{equation}
However $d_0$ does not have a well defined entropy: the localization
at any particular $x=n+\frac{1}{2}$ contributes with $-\infty$ to the
entropy (an ultraviolet divergence) but the freedom to choose any $n$
adds a $+\infty$ (an infrared divergence). This refers both to the
position and the momentum space entropies since they are equal for
$d_0$ ($d_0$ is essentially equal to its Fourier transform).

The function $\psi_0(x)$ is a regularized version of $d_0$ in the
infrared sector (by using $e^{-\pi a^2 x^2}$ instead of $1$) and in
the ultraviolet sector (by using $e^{-\pi x^2/a^2}$ instead of
$\delta(x)$). Remarkably the actual value of the entropy depends on
the concrete choice of the regularization, i.e. there is no natural
value to be assigned to the entropy of $d_0$.

In order to analyze this let us consider a more general regularization
\begin{eqnarray}
\psi(x) &=& 
\aalpha(ax)\sum_{n\in \mathbb{Z}}
(-1)^n \abeta((x-n-\thalf)/a)
\nonumber \\
&=&
\sum_{n\in \mathbb{Z}}
(-1)^n \aalpha( a(n+\thalf))
\abeta((x-n-\thalf)/a) \,. \nonumber \\
\label{eq:3}
\end{eqnarray}
(As always, the equality refers to $a\to 0^+$.) Here $\aalpha(x)$ and
$\abeta(x)$ are two even smooth functions of rapid fall at infinity
which we assume to be normalized. In this case $\psi(x)$ is also
normalized and it will be instructive to show this explicitly. Because
the $\abeta(x)$ falls rapidly at infinity the regularized deltas do
not overlap for small $a$, thus
\begin{eqnarray}
\int dx|\psi(x)|^2 &=&
\int dx 
\sum_{n\in \mathbb{Z}} 
|\aalpha( a(n+\thalf))|^2
|\abeta((x-n-\thalf)/a)|^2
\nonumber \\
&=&
\sum_{n\in \mathbb{Z}} 
|\aalpha( a(n+\thalf))|^2
\int dx 
|\abeta((x-n-\thalf)/a)|^2
\nonumber \\
&=&
a\sum_{n\in \mathbb{Z}} 
|\aalpha( a(n+\thalf))|^2
\int dx 
|\abeta(x)|^2
\nonumber \\
&=& \int dy 
|\aalpha(y)|^2
\int dx 
|\abeta(x)|^2 
\nonumber \\
&=& 1\,.
\label{eq:4}
\end{eqnarray}
A similar calculation for the entropy in position space gives
\begin{equation}
S_x(\psi)= S_x(\aalpha)+S_x(\abeta) \,.
\end{equation}
In order to compute the momentum space entropy, we will use from now
units $2\pi\hbar=1$ which give simpler formulas (in particular
$S_k(\psi)=S_x(\tilde\psi)$). Note that $S_k(\psi)$ has been defined
so that it is numerically independent of $\hbar$ (for fixed
$\psi(x)$). Using Poisson's summation formula
$\sum_n\phi(n)=\sum_n\tilde\phi(n)$, one easily obtains
\begin{eqnarray}
\tilde\psi(k) &=& 
-i\tilde\abeta(ak)\sum_{n\in \mathbb{Z}}
(-1)^n \tilde\aalpha((k-n-\thalf)/a)
\nonumber \\
&=&
-i\sum_{n\in \mathbb{Z}}
(-1)^n \tilde\abeta( a(n+\thalf))
\tilde\aalpha((k-n-\frac{1}{2})/a) \,. \nonumber \\
\end{eqnarray}
This expression is formally identical to $\psi(x)$ in (\ref{eq:3})
replacing $\aalpha$ and $\abeta$ by $\tilde\abeta$ and $\tilde\aalpha$
and so
\begin{equation}
S_k(\psi)= S_x(\tilde\aalpha)+S_x(\tilde\abeta)
= S_k(\aalpha)+S_k(\abeta) \,.
\end{equation}
This implies for the phase space entropy
\begin{equation}
S(\psi)= S(\aalpha)+S(\abeta) \,.
\end{equation}
The minimum is then obtained by choosing Gaussians as the
regularizing functions $\aalpha$ and $\abeta$, as in $\psi_0(x)$.

{\bf 5.} We will now study maximally localized (minimum uncertainty)
states in other subspaces of $L^2(\mathbb{R})$. To this end let us
consider generalizations of the distribution $d_0(x)$. These more
general distributions $u(x)$ will be composed of a set of
well-separated Dirac deltas with different amplitudes and enjoying
some periodicity properties. After suitable infrared and ultraviolet
regularization we can then use the same procedure described above to
compute the entropy. Specifically, we assume
\begin{eqnarray}
&& u(x)= \sum_{k=1}^N b_k\delta(x-x_k)\quad\text{for}\quad 0\le x < r\,,
\nonumber \\
&& 0\le x_1 < \cdots < x_N < r \,,
\nonumber \\
&& |u(x+r)|= |u(x)| \,.
\label{eq:2}
\end{eqnarray}
Assuming that no $b_k$ vanishes, $u(x)$ is the superposition of $N$
series, each series $k=1,\dots,N$ being composed of equidistant deltas
of strength $|b_k|^2$. The period $r$ is common to the $N$ series. In
principle one could consider a larger family of distributions by
taking linear combinations of $u$'s, $\sum_i u_i(x)$ with different
periods $r_i$, however, in order to guarantee that the deltas are
well-separated (and this is essential for being able to compute the
norm and the entropy) the ratios $r_i/r_j$ must be rational numbers.
In this case we are back within the class of distributions defined in
(\ref{eq:2}) with $r$ equal to a common multiple of the $r_i$.

Next we regularize $u(x)$ with normalized functions $\aalpha(x)$ and
$\abeta(x)$ as in (\ref{eq:3}). This amounts to make a convolution of
$u(x)$ with $\abeta(x/a)$ and multiply by $\aalpha(ax)$ (in any
order).  The norm of the resulting function $\psi(x)$ can be computed
along the lines of (\ref{eq:4}) (working with each series of deltas
separately).  This gives
\begin{equation}
\int dx |\psi(x)|^2 =
r^{-1}\sum_{k=1}^N |b_k|^2 \,.
\end{equation}
The norm decreases for increasing $r$ (for fixed $b_k$) since less
strength falls under the profile $\aalpha(ax)$. The entropy in
position space can also be computed (exploiting again that the deltas
are well-separated as $a$ goes to zero) and this gives
\begin{equation}
S_x(\psi) = S_x(\aalpha)+S_x(\abeta)+ S(b)-\log(r)\,,
\label{eq:5a}
\end{equation}
where
\begin{equation}
S(b) = -\sum_{k=1}^N \rho_k\log\rho_k\,,\quad
\rho_k= |b_k|^2/\sum_{j=1}^N|b_j|^2 \,.
\end{equation}

Eq.~(\ref{eq:5a}) nicely shows how the various structures combined in
$\psi(x)$ contribute additively to the entropy. $S(b)$ corresponds to
an entropy for the mixing of the various series in $u(x)$. If the
number of series is small, or more precisely if the strength is
concentrated in a few series, this mixing entropy will decrease.
However, the term $-\log(r)$ indicates that what really matters for
the entropy of $\psi(x)$ is the effective number of series per unit
length. This is sensible since any $u(x)$ with $N$ series and period
$r$ can also be regarded as having $nN$ series and period $nr$ for
$n= 2,3,\ldots$.

Assuming that $\tilde u(x)$ belongs to the same class (\ref{eq:2}) for
some $\tilde b_k$, $\tilde N$ and $\tilde r$, the total phase space
entropy becomes
\begin{equation}
S(\psi) = S(\aalpha)+S(\abeta)+ S(b)+S(\tilde b)-\log(r\tilde r)\,.
\label{eq:5}
\end{equation}
As a consequence the optimum small and large scale profiles $\abeta$
and $\aalpha$ are Gaussians, and we will assume that in what follows.

We note that similar relations are obtained for the $p$-norms, namely
\begin{equation}
\int dx |\psi(x)|^p = r^{-1}\sum_{k=1}^N |b_k|^p \int dy
|\aalpha(y)|^p \int dx |\abeta(x)|^p \,.
\label{eq:6b}
\end{equation}

{\bf 6.} The Fourier transform operator naturally decomposes
$L^2(\mathbb R)$ into the four invariant subspaces corresponding to
its four eigenvalues $\lambda= \pm 1,\pm i$. We want to find maximally
localized states in each such subspace. The Gaussians belong to the
subspace $\lambda= 1$ so they are also the minimizer in this
subspace. Our best guess for antisymmetric functions $\psi_0(x)$
belongs to the subspace $\lambda=-i$, so this is also our best guess
in this subspace. It remains to consider the cases $\lambda=-1$ and
$\lambda=i$. Brute force numerical minimization suggests that the
minimizers in these two cases belong to the class (\ref{eq:2}) with
Gaussian regularization. In order to study this further let us
introduce the set of distributions
\begin{equation}
\phi(x,r,\alpha,\beta)=
\sum_{n\in\mathbb Z}e^{-2\pi i\beta n}\delta(x-r(n+\alpha))\,,
\end{equation}
for non vanishing real $r$ and real $\alpha$ and $\beta$. Not all
these functions are independent since
\begin{eqnarray}
\phi(x,r,\alpha,\beta) &=&
\phi(x,-r,-\alpha,-\beta) =
\phi(x,r,\alpha,\beta+1)
\nonumber \\
&=&
e^{-2\pi i\beta}\phi(x,r,\alpha+1,\beta)
\end{eqnarray}
and all different results are covered by $0<r$, $0\le\alpha,\beta <
1$. Under Fourier transform one finds
\begin{equation}
\tilde\phi(x,r,\alpha,\beta)= r^{-1} e^{2\pi
i\alpha\beta}\phi(x,r^{-1},-\beta,\alpha) \,.
\end{equation}
Therefore, the distributions $\phi(x,r,\alpha,\beta)$ and their
Fourier transform belong to the class in (\ref{eq:2}) with $N=1$.

In order to minimize the mixing entropy $S(b)$ we start by looking for
solutions of the form $\tilde\phi=\lambda\phi$. The unique solutions
are $\phi(x,1,0,0)$ with $\lambda=1$ and
$\phi(x,1,\frac{1}{2},\frac{1}{2})$ with $\lambda=-i$ both with
entropy $1-\log2$ after Gaussian regularization (the latter solution
is just $d_0$).

The next simplest guess is to project a single distribution $\phi$ on
the subspaces $\lambda=-1$ or $\lambda=i$. If $F$ denotes the Fourier
transform operator, the projector on the subspace $\lambda$ is
$P_\lambda=(1+\lambda^{-1}F+\lambda^{-2}F^2+\lambda^{-3}F^3)/4$. Thus
a combination of no more than four $\phi$'s is sufficient to yield a
distribution of the type $\tilde u(x)=\lambda u(x)$. A further
simplification can be achieved as follows. All functions with
$\lambda=-1$ are necessarily even (symmetric), so for these functions
$P_{\lambda=-1}=(1-F)/2$ and only two $\phi$'s are involved after
projection. Similarly, the functions $\lambda=i$ are odd and for these
functions $P_{\lambda=i}=(1-iF)/2$.  In both cases we will seek a
minimum of this form, i.e. projections of a single even or odd $\phi$.
Of course it is not guaranteed that the true minimum is obtained in
this way.

Two remarks should be made. First, the projection $P_\lambda\phi$
involves periods $r$ and $r^{-1}$. This combination is acceptable only
if $r/r^{-1}= r^2$ is a rational number. We will take $r=\sqrt{q/p}$,
where the positive integers $q$ and $p$ have no common divisor.
Second, although each $\phi$ contains a single series of deltas, the
combination of two $\phi$'s may have many more than two series of
deltas. For instance $u(x)=\phi(x,1/2,0,0)+\phi(x,2,0,0)$ has period
$r=2$ and $N=4$ with $x_k=0,1/2,1,3/2$ and $b_k=2,1,1,1$. In practice
$u(x)$ will be composed of two $\phi$'s with periods $r=\sqrt{q/p}$
and $r^{-1}$. In order to disentangle the different series contained
in $u(x)$ it will be necessary to bring all the $\phi$'s to a common
period. To this end the following identity is useful
\begin{equation}
\phi(x,r,\alpha,\beta)=
\sum_{k=0}^{n-1} e^{-2\pi i\beta k}
\phi(x,nr,\frac{k+\alpha}{n},n\beta)\,,
\end{equation}
where $n$ is any positive integer. In our case the smallest common
period is $r^\prime=pr=qr^{-1}=\sqrt{qp}$.

{\sl Case $\lambda=-1$}. There are three classes of single
distributions $\phi$ which are even under $x\to -x$, namely
$\phi(x,r,0,0)$, $\phi(x,r,0,1/2)$ and $\phi(x,r,1/2,0)$, and
arbitrary period $r$. The case $(1/2,0)$ needs not be considered since
it will be generated after projection of the case $(0,1/2)$.

Projection of $\phi(x,r,0,0)$ on the $\lambda=-1$ subspace gives
\begin{equation}
u(x)= \phi(x,r,0,0)-r^{-1}\phi(x,r^{-1},0,0)\,.
\end{equation}
In this case we find a total of $N=q+p-1$ series of period
$r^\prime=\sqrt{qp}$. (Actually $\phi(x,r,0,0)$ gives $p$ series of
period $r^\prime$ and $\phi(x,r^{-1},0,0)$ gives $q$ series of period
$r^\prime$ but one of the series is common to both functions.) Of
these, $p-1$ series have amplitude $b_k=1$, $q-1$ series have
amplitude $b_k=-r^{-1}$ and there is a single series with amplitude
$1-r^{-1}$.  A numerical survey shows that the best case corresponds
to $q=1$ and $p=2$ (or vice versa). This corresponds to a distribution
$u(x)$ composed of only two series of period $r^\prime=\sqrt{2}$ and
weights $\rho_1=(2-\sqrt{2})/4$ an $\rho_2=(2+\sqrt{2})/4$. The mixing
entropy is $S(b)-\log(r^\prime) = \log
2+\frac{1}{\sqrt{2}}\log(\sqrt{2}-1)$. The total phase space entropy
(i.e., adding the Gaussian contribution $1-\log2$ and an overall
factor of two to account for the momentum space entropy) is thus
\begin{equation}
S_{\lambda=-1}= 2+\sqrt{2}\log(\sqrt{2}-1) = 0.753550 \,
\end{equation}
corresponding to the distribution
\begin{equation}
u(x)= (1-\sqrt{2}) \phi(x,\sqrt{2},0,0) + \phi(x,\sqrt{2},1/2,0)\,.
\label{eq:33}
\end{equation}
This is our best guess for the subspace $\lambda=-1$.

The other possibility in this subspace is to project
$\phi(x,r,0,1/2)$.  The analysis is slightly more involved in this
case. If $p$ is odd the number of series is $p+q$ and the total
entropy is $S=2$ for any choice of $p$ and $q$. If $p$ is even the
number of series is $q+p-1$ and the entropy depends of the concrete
values of $q$ and $p$. The best choice corresponds to $q=1$, $p=2$ and
yields the same distribution $u(x)$ in (\ref{eq:33}).

{\sl Case $\lambda=+i$}. The only possibility of an odd $\phi$ is
$\phi(x,r,1/2,1/2)$. This gives
\begin{equation}
u(x)= \phi(x,r,1/2,1/2)-r^{-1}\phi(x,r^{-1},1/2,1/2)\,.
\label{eq:6}
\end{equation}
If $p$ or $q$ is even the number of series is $q+p$ and for all
choices of $q$ and $p$ the total entropy is 2. If $q$ and $p$ are odd
the number of series is $q+p-1$. The best choice corresponds to $q=3$
and $p=1$ (or vice versa) with a total phase space entropy
\begin{equation}
S_{\lambda=+i}= 2-\frac{2}{\sqrt{3}}\log(\sqrt{3}+1) = 0.839465
\end{equation}
This is our best guess for the subspace $\lambda=+i$.

These results are well below the guesses based on the lowest harmonic
oscillator states which yield $S= 1.15934$ for $\lambda=-1$ and $S=
1.38155$ for $\lambda=i$.

By using (\ref{eq:6b}) these results extend immediately to relations
for the norm of the Fourier transform operator from
$L^p({\mathbb{R}}^d)$ into $L^q({\mathbb{R}}^d)$, restricted to the
proper subspaces $\lambda=\pm 1,\pm i$,

\begin{acknowledgments}
This work is supported in part by funds provided by the Spanish DGI
and FEDER founds with grant no. BFM2002-03218, and Junta de
Andaluc\'{\i}a grant no. FQM-225.
\end{acknowledgments}


\end{document}